# Size effect on polarization bremsstrahlung emission from xenon clusters

Yu.S. Doronin, A.A. Tkachenko, V.L. Vakula, G.V. Kamarchuk

B. Verkin Institute for Low Temperature Physics and Engineering

of the National Academy of Sciences of Ukraine,

Kharkiv 61103, Ukraine

E-mail: doronin@ilt.kharkov.ua

**Abstract**

We measured the spectral distribution of the absolute differential cross section of both ordinary and polarization ultra-soft X-ray bremsstrahlung for 0.7 keV electrons scattered on substrate-free nanoclusters of xenon. Clusters were produced in a supersonic gas jet, expanding adiabatically into a vacuum. An original method based on absolute measurements of the intensity of the atom and cluster emission in the vacuum ultraviolet and ultra-soft X-ray spectral regions was used to determine the cluster density in the scattering area. The bremsstrahlung arising from scattering of electrons on clusters had a polarization component which dominated the differential cross section. For the first time, cluster size effect on the formation of the polarization bremsstrahlung was found for xenon.

1. **Introduction**

Nanometer-sized aggregates of matter - clusters - are attractive objects of study in various fields of science, such as physics, chemistry, medicine, and materials science. Clusters are vital to many processes in the atmosphere and space. Controlling the particle size, the crystallographic structure and morphology of nanoparticles is critically important today, both from fundamental and industrial perspectives. Supersonic free jets of rare-gas clusters [1-3] make it possible to isolate and characterize molecules and complexes at very low temperatures, inaccessible to other techniques.

The interaction of clusters of rare gases with electrons is an important area of research because it provides a tool for a better understanding cluster properties. Collisions of electrons with clusters of different sizes are accompanied by complex interactions that cause the emission of radiation over a wide spectral range, including the continuous X-ray emission known as bremsstrahlung [4-7].

Bremsstrahlung (BS) [8,9], arising from the scattering of electrons by atoms, molecules, or clusters, is formed by two mechanisms. One of them implies that the emission of photons is caused by an incident electron when it is decelerated in the static field of a target particle, which is ordinary bremsstrahlung (OBS). According to the other mechanism, continuous-spectrum photons are emitted by the electrons of the target particle due to their dynamic polarization, which is the case of polarization bremsstrahlung (PBS). In addition to differences in the formation mechanisms, OBS and PBS differ in several other key aspects: they depend differently on the incident energy, have different angular distributions, and while OBS emission is formed when the incident electron is close to the target, PBS emission is formed at a greater distance away [10

Experimental and theoretical studies of OBS and PBS have received much attention in recent decades, which has stimulated the creation of new experimental techniques and the development of theoretical analysis methods. The direct observation of PBS was first carried out in the ultra-soft X-ray (USX) region at photon energies of 70–220 eV during the scattering of electrons with energies of 0.6 keV on Xe atoms [11]. Polarization bremsstrahlung manifested itself as a broad band with a structure similar to the 'giant' resonance in the Xe atom photoabsorption. Sal Portillo and K. A. Quarles [12], in their measurement of the absolute differential cross sections of bremsstrahlung during the bombardment of free Xe, Kr, Ar, and Ne atoms by electrons with energies of 28 and 50 keV for the first time determined the contribution of polarization bremsstrahlung to the photon spectrum from electron bremsstrahlung. In their review [13], the authors discussed double differential cross-section experiments for electron bremsstrahlung from free gas atoms and thin-film targets for 100 keV or less electron energies. They compared the cross-section ratios for different target atoms within two theoretical models: ordinary bremsstrahlung and total bremsstrahlung calculated in the stripping approximation.

Further experimentally obtained BS cross sections for different energies of electrons scattered on thin targets made of C, Al, Te, Ta, and Au [14] showed that a significant discrepancy between the theoretical and experimental results appears with decreasing photon energy of BS. The absolute BS differential cross sections for clusters are of great interest because they demonstrate a significant discrepancy between the measured and calculated BS differential cross sections due to the pronounced PBS contribution to their BS spectra.

The present work aims to study in detail the process of formation of OBS and PBS induced by the scattering of intermediate-energy electrons on xenon clusters and find the absolute double differential BS cross sections for different cluster sizes. Obtaining such data may be necessary for the development of the theory of bremsstrahlung covering both atoms and clusters.

## 2. Experiment

The experimental method used in this work was described in detail in Ref. [10,15]. The experimental setup consisted of an X-ray tube with a supersonic gas jet used as an anode and an X-ray spectrometer. The gas jet technique for producing rare-gas atom-cluster targets offers several advantages:

• By adjusting the gas parameters at the nozzle entrance, the target type can be varied from atoms to clusters, and the size and density of clusters in the jet can be accurately controlled.

• The method allows the detection of the generated radiation directly within the X-ray spectrometer, eliminating the need for optical windows and differential evacuation systems.

• A well-defined boundary of the cluster jet in a vacuum allows the electron gun to be placed near the jet, which makes it possible to excite nanoclusters by an electron beam with a high current density at relatively low energies ranging from 200 to 2000 eV.

A supersonic conical nozzle formed a jet that flowed adiabatically into a vacuum chamber. A beam of clusters and atoms, depending on the conditions of the experiment, was produced at the nozzle outlet. We varied the cluster size from 400 to 12000 atoms per cluster by varying the gas temperature (180 to 500 K) and pressure (0.03 to 0.12 MPa) at the nozzle inlet. An electron beam crossed the xenon jet at a distance of 10 mm from the nozzle outlet. The electron energy was 0.3–1.0 keV, while the electron current was kept constant at 20 mA. When electrons interact with an atomic or cluster target, the resulting radiation passes through the entrance slit of an X-ray spectrometer and is detected by a proportional counter filled with methane at a pressure of $1.5 \times 10^4$ Pa. The counter had a window made of a nitrocellulose film. Differential BS spectra were recorded with a solid radiation collection angle of $1.7 \times 10^{-3}$ sr and a spectral resolution of 1 Å. The angle between the incident electron beam and the detected photons was set to 97°. Systematic studies show that, under constant experimental conditions, the BS intensities for atomic and cluster Xe jet remain stable over time, with an experimental error margin of 10%. The experiments were conducted in the single-collision regime, as evidenced by the observed linear relationship between BS intensity and electron beam current across the 5–30 mA range.

The spectral distribution of the flux density of the bremsstrahlung generated in the photon energy range of 70–200 eV at electron scattering on a xenon cluster jet was determined in absolute units by a method [15] developed earlier. To determine the absolute intensity of the bremsstrahlung $I^{BS}$, a spectral dependence of the absolute sensitivity of the X-ray spectral equipment was determined using a calibrated supersonic argon jet. A silicon photodiode SXUV - 100 with known spectral sensitivity in absolute units for the 5-50 nm wavelength range was used as a radiation detector. The spectral distribution of BS in absolute units for 0.7 keV electrons scattered by xenon clusters of two average sizes $<N> = 400$ and $<N> = 12000$ atoms/cluster is shown in Fig. 1.

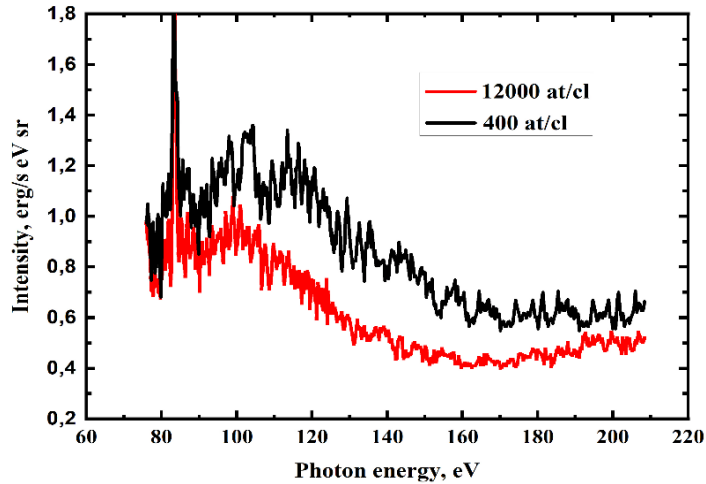

Fig. 1. Spectral distribution of BS in absolute units for 0.7 keV electrons scattered by xenon clusters of two average sizes $<N> = 400$ and $<N> = 12000$ atoms/cluster.

The measured spectra are a superposition of the PBS and OBS spectra. Similar behaviour of the BS spectral intensity was observed for all sizes of xenon clusters studied in this work. The BS dependence is defined by a broad maximum extending into the high photon-energy region. The intensity maximum is at the energy ~105 eV. In the 80-90 eV region, the superposition of narrow characteristic lines corresponding to the $4p^54d^9 \rightarrow 4d^8$ transition is noticeable [16]. These results are a baseline for finding differential BS cross sections in absolute units.

The methodology for correctly extracting the accurate PBS profile from the whole BS was described in Ref. [10]. The double differential total BS cross section in barn/eVsr units was calculated using the following relation:

$$\frac{d^2\sigma^{BS}}{d\omega d\Omega} = \frac{I^{BS}(\hbar\omega)}{n_{cl} n_e v \hbar\omega} \cdot \frac{1}{A}, \quad (1)$$

Here, $I^{BS}(\hbar\omega)$ [erg/s eV] is the BS intensity at the photon energy $\hbar\omega$ [eV] in the solid angle $d\Omega$; $n_{cl}$ and $n_e$ [cm$^{-3}$] are the densities of clusters and electrons, respectively; $v$ [cm/s] is the electron velocity; and $A$ [cm$^3$] is the excitation region volume.

The density of xenon clusters $n_{cl}$ at the intersection of the electron beam with the supersonic gas jet was determined by the technique tested on argon clusters [17]. It is based on the fact that the temperature dependence of the intensities of the ion and resonance lines of rare atoms is directly proportional to the amount of uncondensed matter in the supersonic jet. Having measured the intensity $\Phi(\lambda)$ of the XeI resonance line ($\lambda$ = 147 nm) in absolute units and knowing its absolute emission cross sections $\sigma(\lambda)$ [18], we can determine the atomic matrix density of the jet under different flow conditions using the following relation:

$$n = \frac{1}{\sigma(\lambda) l \frac{i}{e}} \frac{4\pi}{\Omega} \Phi(\lambda), \quad (2)$$

$n$ - target gas density [cm$^{-3}$],

$\sigma(\lambda)$ – absolute emission cross sections [cm$^2$] for XeI line ($\lambda$ = 147 nm),

$l$ - length of an observed electron beam [cm],

$\Omega$ - solid angle of observation [sr],

$i$ - electron beam current [A],

$e$ - electron charge [C],

$\Phi(\lambda)$ – absolute radiant flux [photon/s] for $\lambda$ = 147 nm.

The contribution of cascade processes and self-absorption to the intensity of the Xe ($^3P_1$) resonance line at 147 nm [5] for the excitation electron energy of about 1000 eV is less than 10% so that we can speak of a direct proportionality of its intensity to the amount of uncondensed matter in the jet. The presence of a strong inverse correlation ($r \approx -1$) between the temperature dependences of the intensity of the cluster molecular emission at 176 nm and the resonance line at 147 nm allowed us to determine the absolute values of the condensate fraction $f_c$ [cm$^{-3}$] as the total number of jet atoms participating in the generation of Xe clusters at different gas flow temperatures. Furthermore, using the methodology described in Ref. [19, 20], we calculated the average cluster size $<N>$ [at/cl] (see Fig. 2) for our supersonic jet flow parameters and the density $n_{cl} = f_c /<N>$ [cm$^{-3}$] of Xe clusters in the jet as a function of their size (see Fig. 3). The uncertainty in the determination of the average size was 15%.

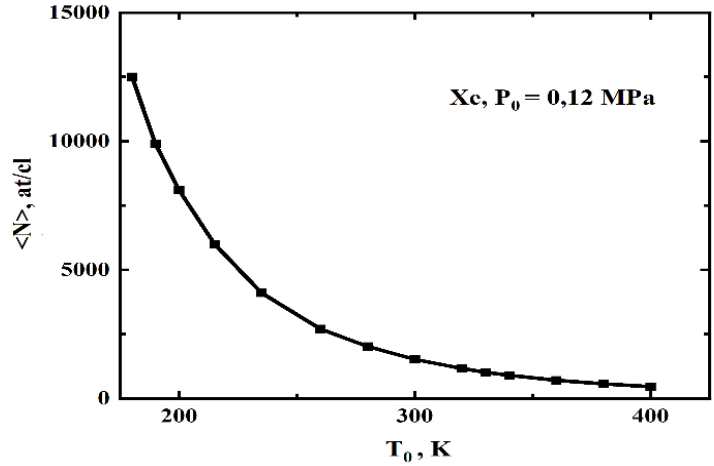

Fig. 2. Average cluster size $<N>$ on the temperature $T_0$ and pressure $P_0$.

The dependence of the density of clusters per unit volume shown in Fig. 3 reflects the presence of different stages of cluster formation in a supersonic jet. The first stage is the formation of condensate nuclei in the jet up to $<N>$ = 1500 at/cl due to three-particle collisions and their further growth by coalescence and adhesion of atoms to the surface of clusters.

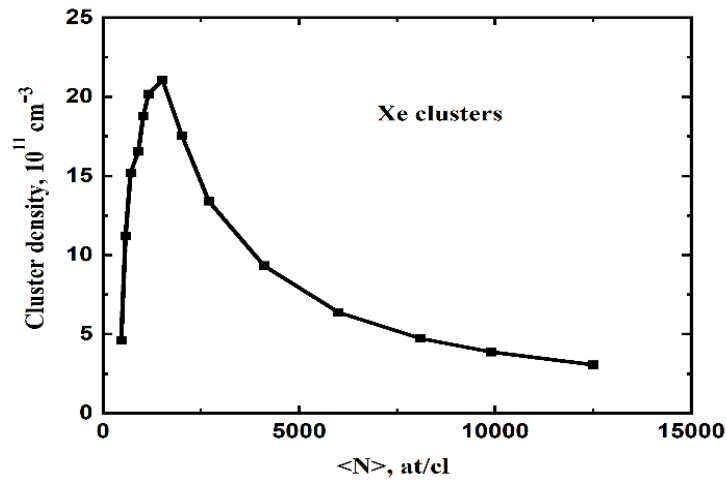

Fig. 3. Density of clusters in the studied region of the jet.

In the second stage, above 1500 at/cl, the coagulation process (cluster sticking) dominates, and the growth of the condensate fraction in the jet stops.

The double BS differential cross sections for xenon clusters calculated with Eqs. (1) are presented in Fig. 4 as $\omega d^2\sigma(\omega)/d\omega d\Omega$ vs. photon energy $\hbar\omega$. The uncertainty in the determination of the differential cross sections was 40 %. It should be noted that in the above coordinates, the OBS differential cross section is constant and does not affect the PBS profile [10]. For comparison, Fig. 4 also shows a similar dependence for xenon atoms.

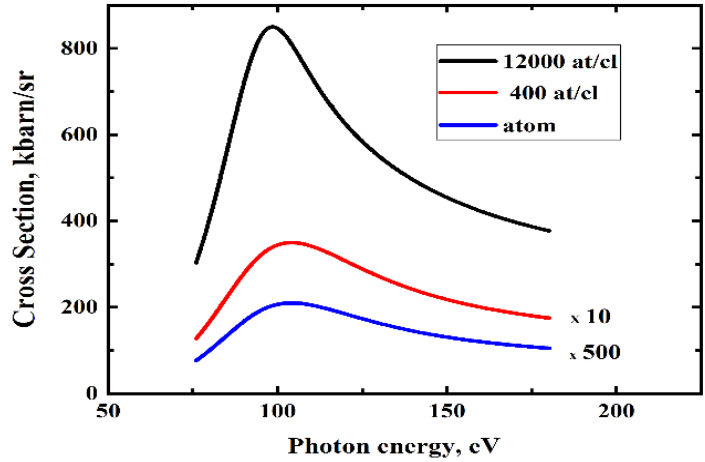

Fig. 4. Absolute differential BS cross section for 0.7 keV electrons scattered on Xe atoms and Xe clusters of different sizes in the photon energy range of 70-200 eV.

The spectrum exhibits a well-defined broad maximum in the photon energy range of 80–155 eV, associated with the polarization bremsstrahlung [12, 13]. The BS cross section shape does not virtually change, while its numerical value increases significantly with increasing cluster size. Analysis of the data in Fig. 4 makes it possible to estimate the contribution of PBS to the total BS for xenon clusters of different sizes.

Figure 5 shows dependencies of absolute differential cross section of total bremsstrahlung (OBS+PBS) and ordinary bremsstrahlung (OBS) on the square radius of xenon clusters.

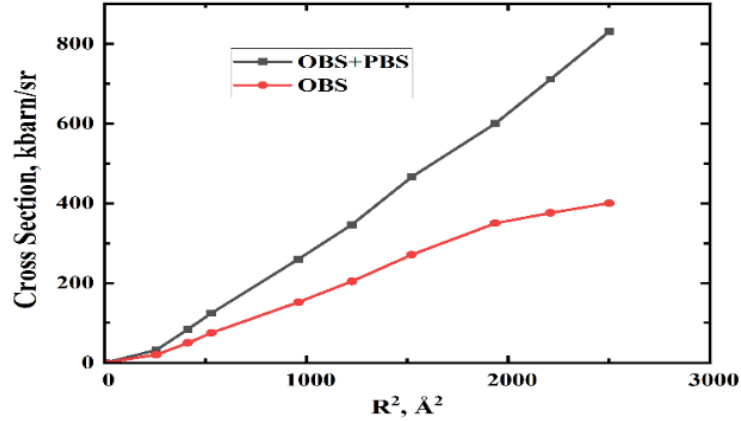

Fig. 5. Dependencies of the absolute differential cross-section of total bremsstrahlung (OBS+PBS) and ordinary bremsstrahlung (OBS) on the square radius of xenon clusters.

It can be seen from Fig. 5 that differential cross sections, both (OBS+PBS) and OBS, increase with increasing cluster size. For clusters with 3000 or more atoms per cluster (square radii of 970 Å$^2$ and more), the contribution of the polarization component to the differential cross section is the dominating one. That may be due to multiparticle interactions and interference of the contributions of atoms within clusters to the formation of PBS. The results show that cluster cooperative phenomena significantly influence the main characteristics of bremsstrahlung.

### 3. Conclusions

The experimental methodology used in this work allowed us to obtain for the first time the spectral distribution of the absolute differential cross section of the total BS (including PBS) from a scattering of electrons of intermediate energies (0.7 keV) on free Xe clusters in the photon energy range of 70 - 200 eV. The study shows that the polarization component of bremsstrahlung significantly dominates the differential cross section for electron scattering on xenon clusters. That is, the size and structure of clusters play a significant role in the emission characteristics of PBS.

An original method was applied based on absolute measurements of the emission intensity of atoms and clusters in the vacuum ultraviolet and ultrasoft X-ray spectral regions. This technique was crucial to accurately determine the density of clusters in the scattering region, which is necessary to understand bremsstrahlung processes.

The study opens new avenues for fundamental research in atomic and cluster physics. Providing a deeper understanding of the mechanisms behind PBS encourages further exploration of electron interactions with different materials, which can lead to discoveries in quantum mechanics and condensed matter physics.

In summary, the practical implications of this research extend from advancements in technology and material science to fundamental physics, highlighting the importance of understanding polarization bremsstrahlung in various applications.

**Acknowledgments**

The authors are grateful to Dr. O. Konotop for his assistance in preparing the paper.